\documentclass[conference]{IEEEtran}
\usepackage{cite}
\usepackage{amsmath,amssymb,amsfonts}
\usepackage{algorithmic}
\usepackage{graphicx}
\usepackage{textcomp}
\usepackage{xcolor}
\usepackage[hyphens]{url}

\usepackage{color,xcolor}
\usepackage{xspace}
\usepackage{multirow}
\usepackage{booktabs}
\usepackage{lipsum}
\usepackage{soulutf8}
\usepackage{diagbox}
\usepackage{physics}
\usepackage[utf8]{inputenc}
\usepackage[T1,OT1]{fontenc}
\usepackage{colortbl}

\usepackage{enumitem}
\usepackage{braket}
\usepackage{calc}

\usepackage{amsmath}
\usepackage{amsfonts}
\usepackage{url}

\usepackage{afterpage}
\usepackage{cancel}

\usepackage{algorithmic}
\usepackage{algorithm}
\usepackage{setspace}

\usepackage{pifont}

\usepackage{titlesec}

\usepackage{hyperref}
\hypersetup{
    colorlinks=true,
    linkcolor=magenta,
    filecolor=magenta,      
    urlcolor=blue,
}

\usepackage{xcolor}
\usepackage{soul}

\definecolor{citecolor}{RGB}{34,139,34}
\definecolor{mydarkblue}{rgb}{0,0.08,1}
\definecolor{mydarkgreen}{rgb}{0.02,0.6,0.02}
\definecolor{mydarkred}{rgb}{0.8,0.02,0.02}
\definecolor{mydarkorange}{rgb}{0.40,0.2,0.02}
\definecolor{mypurple}{RGB}{111,0,255}
\definecolor{myred}{rgb}{1.0,0.0,0.0}
\definecolor{mygold}{rgb}{0.75,0.6,0.12}
\definecolor{myblue}{rgb}{0,0.2,0.8}
\definecolor{mydarkgray}{rgb}{0.,0.2,0.2}

\definecolor{lightred}{RGB}{255,235,235}
\definecolor{lightgreen}{RGB}{235,255,235}
\definecolor{lightblue}{RGB}{235,235,255}
\definecolor{lightcyan}{RGB}{235,255,255}
\definecolor{lightmagenta}{RGB}{255,235,255}
\definecolor{lightyellow}{RGB}{255,255,235}

\definecolor{qxkcolor}{RGB}{215,235,255}
\definecolor{softmaxcolor}{RGB}{230,235,255}
\definecolor{probxvcolor}{RGB}{255,255,235}

\definecolor{topkcolor}{RGB}{255,235,235}
\definecolor{zecolor}{RGB}{255,255,235}
\definecolor{dynacolor}{RGB}{235,255,255}

\definecolor{reviewcolor}{RGB}{0,0,200}

\renewcommand\footnotemark{}

\newcommand\blfootnote[1]{%
  \begingroup
  \renewcommand\thefootnote{}\footnote{#1}%
  \addtocounter{footnote}{-1}%
  \endgroup
}

\newcommand{\name}{DGR\xspace}
\newcommand{\mwpm}{MWPM\xspace}

\newcommand{\mwpmfull}{Minimum-Weight-Perfect-Matching\xspace}

\newcommand{\x}{$\times$\xspace}

\newcounter{rlabelno}

\definecolor{color1}{rgb}{1, 1, 0.9}
\definecolor{color2}{rgb}{1, 0.9, 1}
\definecolor{color3}{rgb}{0.9, 1, 1}

\definecolor{color4}{rgb}{1, 0.9, 0.9}
\definecolor{color5}{rgb}{0.9, 0.9, 1}
\definecolor{color6}{rgb}{0.9, 1, 0.9}

\definecolor{color7}{rgb}{0.8, 0.9, 1}
\definecolor{color8}{rgb}{0.9, 1, 0.8}
\definecolor{color9}{rgb}{1, 0.8, 0.9}

\newcolumntype{a}{>{\columncolor{color1}}c}
\newcolumntype{b}{>{\columncolor{color2}}c}
\newcolumntype{d}{>{\columncolor{color3}}c}
\newcolumntype{e}{>{\columncolor{color4}}c}
\newcolumntype{f}{>{\columncolor{color5}}c}
\newcolumntype{g}{>{\columncolor{color6}}c}
\newcolumntype{h}{>{\columncolor{color7}}c}
\newcolumntype{i}{>{\columncolor{color8}}c}
\newcolumntype{j}{>{\columncolor{color9}}c}

\def\BibTeX{{\rm B\kern-.05em{\sc i\kern-.025em b}\kern-.08em
    T\kern-.1667em\lower.7ex\hbox{E}\kern-.125emX}}

\title{\name: Tackling Drifted and Correlated Noise in Quantum Error Correction via \underline{D}ecoding \underline{G}raph \underline{R}e-weighting}

\author{
\IEEEauthorblockN{Hanrui Wang*$^{1}$, Pengyu Liu*$^{2}$, Yilian Liu$^{3}$,  Jiaqi Gu$^{4}$, Jonathan Baker$^{5}$, Frederic T. Chong$^{6}$, Song Han$^{1}$}
\IEEEauthorblockA{$^{1}$Massachusetts Institute of Technology $^{2}$Carnegie Mellon University  $^{3}$Cornell University \\ $^{4}$Arizona State University, $^{5}$University of Texas at Austin, $^{6}$University of Chicago}
}

\begin{document}
\maketitle
\thispagestyle{plain}
\pagestyle{plain}

\begin{abstract}

Quantum hardware suffers from high error rates and noise, which makes directly running applications on them ineffective. Quantum Error Correction (QEC) is a critical technique towards fault tolerance which encodes the quantum information distributively in multiple data qubits and uses syndrome qubits to check parity. A classical decoder is needed to repeatedly process the syndrome qubit information and identify the errors. Minimum-Weight-Perfect-Matching (MWPM) is a popular QEC decoder that takes the syndromes as input and finds the matchings between syndromes that infer the errors. However, there are two paramount challenges for MWPM decoders. First, as noise in real quantum systems can drift over time, there is a potential misalignment with the decoding graph's initial weights, leading to a severe performance degradation in the logical error rates. 
Second, while the MWPM decoder addresses independent errors, it falls short when encountering correlated errors typical on real hardware, such as those in the 2Q depolarizing channel. Overlooking such correlations can adversely impact the performance of the decoder.

To tackle these two challenges, we propose DGR, an efficient decoding graph edge re-weighting strategy with no quantum overhead. It leverages the insight that the statistics of matchings across decoding iterations offer rich information about errors on real quantum hardware. By counting the occurrences of edges and edge pairs in decoded matchings, we can statistically estimate the up-to-date probabilities of each edge and the correlations between them. The reweighting process includes two vital steps: alignment re-weighting and correlation re-weighting. The former updates the MWPM weights based on statistics to align with actual noise, and the latter adjusts the weight considering edge correlations. These adaptations pave the way for a more accurate and resilient quantum error correction pipeline.

Extensive evaluations on surface code and honeycomb code under various settings show that DGR reduces the logical error rate by 3.6\x on average-case noise mismatch with exceeding 5000\x improvement under worst-case mismatch. 

\end{abstract}

\section{Introduction}

\begin{figure}
    \centering
    \includegraphics[width=\columnwidth]{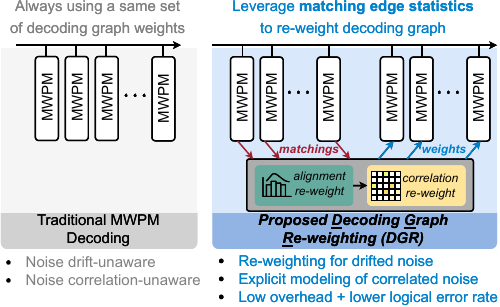}
    \caption{Compare traditional decoding method with our proposed decoding graph re-weighting (DGR).
    DGR leverages decoding statistics to dynamically re-weight the decoding graph for high-fidelity quantum error correction.}
    \label{fig:teaser}
\end{figure}

\blfootnote{*Equal Contributions.}Quantum Computing (QC) has garnered substantial research interest as an emergent computational model designed to address problems previously deemed unsolvable with enhanced efficiency. A multitude of sectors and academic disciplines stand to gain from the potentialities of QC, notably cryptography~\cite{shor1999polynomial}, database search~\cite{grover1996fast}, combinatorial optimization~\cite{farhi2014quantum, liang2023hybrid}, molecular dynamics~\cite{peruzzo2014variational}, and machine learning~\cite{lloyd2013quantum, liang2021can, wang2022quantumnas, wang2022quantumnat, wang2022qoc, wang2022quest, wang2023robuststate, zheng2022sncqa, cheng2022topgen, liang2022pan} applications, etc.

Despite the exciting advancements, the qubits and quantum gates on current quantum machines suffer from high error rates of $10^{-3}$ to $10^{-2}$, preventing us from executing applications that demand significantly lower error rates (below $10^{-10}$)~\cite{lee2021even, kivlichan2020improved, gidney2021factor}.
Therefore, reducing quantum error is a pressing demand to close the gap. Quantum Error Correction (QEC), an essential solution to this challenge, lowers the error rate by integrating redundancy, a process where the information from a single logical qubit is distributed across multiple physical qubits (\textit{data qubits}). 
\textit{Syndrome qubits} then perform iterative checks of the data qubits and provide the parity of them in classical bits, which are then decoded by a classical decoder to infer the error that occurs on data qubits.
By increasing the redundancy, the logical error rate plummets exponentially.

An effective decoder is critical for QEC to lower the error rate. Minimum-Weight-Perfect-Matching (MWPM) is a promising candidate with high accuracy and almost-linear complexity~\cite{fowler2012towards, higgott2023sparse}.
In \mwpm, a \textit{decoding graph} is constructed based on the selected \textit{code and noise model}. Each node in the graph represents a syndrome qubit, and each edge represents an error that can cause a syndrome flip event on its connected nodes. The weight on each edge ($w$) is computed from the error rate of the corresponding error ($p$) as $w=-\log{\frac{p}{1-p}}\approx -\log{p}$.
To deal with the degeneracy between syndrome and errors, i.e., the same syndrome pattern can be caused by different errors, the \mwpm finds the most likely error, which corresponds to the overall minimum weights of the selected error edges. Therefore, a precise noise model is crucial for the high accuracy of the decoder.

However, \mwpm decoders suffer from the \textbf{inaccurate modeling of the real hardware noises} that ignores the temporal drift of the noise and correlation between noises.

\ding{202} \textbf{Noise Drift-Unaware} -- First, the noise on real quantum hardware can suffer significant drifts overtime~\cite{ravi2023navigating} due to various causes such as thermal fluctuations in Josephson Junctions~\cite{gumucs2022calorimetry}, laser detuning in neutral atom and trapped-ion~\cite{day2022limits}, unwanted coupling to two-level-systems (TLS)~\cite{muller2019towards} defects in Transmon qubits~\cite{burnett2019decoherence, schlor2019correlating}, magnetic fluctuation caused by the sun~\cite{mills2022sun}, etc.
The drifted noise creates a \textit{mismatch} between the actual noise characteristics on quantum hardware and the noise model inside the classical \mwpm decoder. 
Though there exist traditional methods for identifying noise in quantum systems, e.g., quantum process tomography~\cite{chuang1997prescription, PhysRevLett.78.390}, randomized benchmarking~\cite{emerson2005scalable, knill2008randomized, magesan2011scalable}, they are time-consuming and cannot be performed simultaneously with QEC in progress.

\ding{203} \textbf{Noise Correlation-Unaware} -- Another issue is the lack of correlation modeling between noises (represented as edges in the decoding graph). Typically, edges within this graph are treated as \textit{independent}. Yet, in many cases, these edges exhibit nontrivial \textit{correlations}. As an illustrative example, consider the widely-referenced X-Z surface code. Here, a Y error in the 1Q depolarizing channel decomposes into two ostensibly independent edges: one X and one Z. However, when a Y error occurs in practice, both edges should be co-selected. Therefore, the upper-bound accuracy of \mwpm is unfortunately compromised due to its neglect of edge correlations.

To overcome the above two issues and augment the existing \mwpm pipeline with more accurate noise modeling, we propose an efficient, non-intrusive, and scalable approach -- Decoding-Graph Re-weighting (\name) (Fig.~\ref{fig:teaser}). 
\name operates based on two pivotal insights. The first recognizes that quantum error correction inherently possesses the ability to \textit{detect} errors, thus allowing the reconstruction of noise characteristics from multiple rounds of QEC. The second insight acknowledges that the method can discern not only independent noise but also the correlations between noises. This correlation information can be harnessed to fine-tune the weights on the graph edges, thereby offering a more robust solution. 

\begin{figure}[t]
    \centering
    \includegraphics[width=\columnwidth]{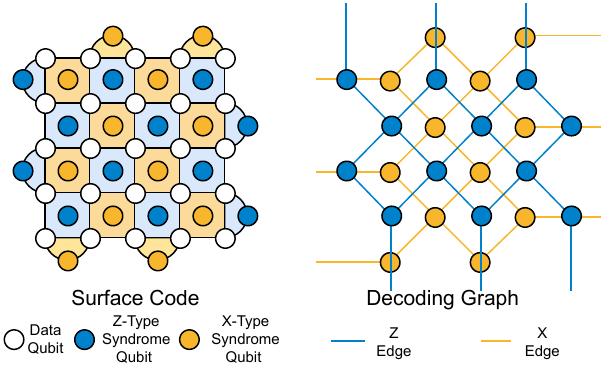}
    \caption{Surface code and its decoding graph.}
    \label{fig:decoding_graph}
\end{figure}

To address the noise drift issue, we propose an \textit{alignment re-weighting} strategy. In QEC, the syndrome extraction process is performed iteratively until the whole quantum program is completed. That provides massive information about the errors on quantum hardware. Take the common standard for real-time decoding \cite{battistel2023real}, 1 $\mu$s per round, 1 million rounds of matchings can be obtained every second. The occurrence of an edge in the matching indicated a predicted error on that edge. Therefore, we first leverage an \textit{occurrence tracer} to trace the frequency of each edge that occurred in the matchings. Given enough rounds, we can accurately estimate the correct weight for every edge. Then, the \mwpm can leverage the updated weights in the later rounds to improve accuracy. This re-weighting is non-intrusive, running in parallel on the classical sides, with low overhead. For instance, with a distance of 5 and a physical error rate of 0.05, 10,000 rounds are enough to have an accurate estimation.

To further consider noise correlation, we propose \textit{correlation re-weighting}. A \textit{correlation tracer} is used to count the frequency of co-occurrence of every edge pair and construct a correlation matrix. Then, we adopt a two-iteration MWPM in this strategy. In the first iteration, we perform a normal \mwpm, and then, according to the edges in the initial matching, we slightly adjust the weights of each edge based on the correlation matrix. We will increase the prob of an edge if its correlated edges are selected in the initial mapping. To give a simple example, if a 1Q X edge is selected, then we will increase the correlated Z edge because the X error may be due to a Y error. Then, we will use the re-weighted graph to perform the second iteration \mwpm and obtain the final matchings. We provide two re-weighter options, one heuristic-based and one neural network (NN)-based, to determine the new weights of the edges according to correlations.
To reduce the overhead, we only perform this re-weighting for difficult syndrome patterns, such as the ones with a relatively large number of syndromes.

\name is a plug-and-play scheme that can be combined with a variety of existing optimizations on \mwpm. More importantly, it incurs \textbf{zero quantum overhead} since there is no additional operation on quantum hardware.

We extensively evaluate \name on over 40 benchmarks with Surface Code and Honeycomb Code under different code distances, error rates, and noise models (phenomenological and circuit-level). 
\name brings more logical error rate reduction with a larger code distance and a lower physical error rate, with 3.6\x and 1.7\x average-case reduction for surface code and honeycomb code. 
For the worst-case mismatch, the reduction for surface code is 695\x on average, up to 7,360\x. 
The correlation re-weighting further reduces the error rate by an average of 1.2\x and up to 1.4\x. 
We also perform sensitivity analysis on the effectiveness of \name and observe its superior capability to recover correct weights even under a 500\x error rate drift. 
The key insight we derive is that, even if the mismatch causes a much higher logical error rate, its absolute value remains much smaller than 1, such that most of the matchings, especially for simple cases, are still correct and can be used to recover the correct weights. 

Our contributions are summarized as follows:
\begin{itemize}[leftmargin=*]
\setlength{\itemindent}{0.1pt}
    \item We show that the mismatch between the noise model in \mwpm and real hardware noise results in a drastic degradation of logical error rates by several orders of magnitude.
    \item To augment the current \mwpm flow with more accurate noise modeling, we propose an efficient, non-intrusive, and scalable approach, DGR, with zero quantum overhead.
    \item We propose alignment re-weighting to trace the frequency of edges from matchings and update edge weights. The process can fully recover the logical error rates to reach similar rates to a weight oracle.
    \item We propose a correlation re-weighting scheme that first traces the correlations between edge pairs based on matchings and then adjusts the edge weights with a simple heuristic or an NN-based re-weighter.
    \item We evaluate \name on surface code and honeycomb code under various settings and achieve 3.6\x and 1.7\x average-case improvement. The worst-case improvement can exceed 5000\x for surface code.
\end{itemize}

\section{Background and Related Work}
\subsection{Surface Code and Honeycomb Code}
The Surface code is a prominent QEC scheme that encodes a logical qubit into a two-dimensional lattice of alternating data and syndrome (parity) qubits. It has a high error threshold and requires only nearest-neighbor connectivity, making it practical for real quantum systems. Errors on data qubits are categorized into a discrete set of Pauli errors - bit-flip (Pauli-X), phase-flip (Pauli-Z), or both~\cite{nielsen2002quantum}. They can be detected by adjacent parity qubits using a stabilizer circuit, which measures a four-qubit operator, leading to the detection of X, Z, or Y (combination of X and Z) errors. The surface code can correct error chains up to length $\lfloor \frac{d-1}{2} \rfloor$. A simpler variant of the original Toric code~\cite{Kitaev_2003}, the `rotated' surface code as in~\ref{fig:decoding_graph} left is often preferred due to its more compact layout, reducing the physical qubits and gate overheads. The $[[d^2, 1, d]]$ stabilizer code has become a prime candidate for near-term fault-tolerant quantum computation. 

Though the surface code has many merits, it requires connectivity of 4. It could be directly mapped to some superconducting hardware~\cite{google2023suppressing} and neutral atom devices~\cite{wang2023qpilot, wang2023fpqac} which means each qubit needs to connect with 4 other qubits, but may be difficult for some others. For example, IBM's newest quantum computer employs a heavy-hexagon lattice, which provides connectivity of only 2.4~\cite{chamberland2020topological}. Such sparse connectivity challenges the design of quantum error correction code. Recently, Honeycomb code~\cite{hastings2021dynamically,haah2022boundaries}, a dynamically generated quantum error correction code, is proposed to perform error correction on quantum computers with heavy-hexagon topology. As a \textit{instantaneous stabilizer} code, it shares many similar features to conventional stabilizer code: in each step, the code can be viewed as stabilizer code, but the stabilizer group changes over time. An advantage of honeycomb code is that its decoding can also be handled with \mwpm~\cite{gidney2022benchmarking,gidney2021fault}. It is worth noticing that the measurement error, single qubit error, and two qubits error will all be correlated edges in the decoding graph. Thus, decoding with an \mwpm decoder is not optimal for honeycomb.

\subsection{Minimum-Weight-Perfect-Matching}

To decode errors from syndromes, researchers have proposed multiple families of decoders such as Lookup Table (LUT) based~\cite{das2022lilliput, tomita2014low, ryan2021realization}, Neural Network (NN) based~\cite{PhysRevLett.119.030501, Breuckmann2018scalableneural, chamberland2022techniques, wang2023transformerqec} and Union-Find (UF) based~\cite{delfosse2020linear, delfosse2021almost, wu2022interpretation, das2022afs}. 
Among various decoders, \mwpmfull decoders~\cite{edmonds1965maximum, dennis2002topological, fowler2012surface, holmes2020nisq} are considered to have a good balance between the decoding accuracy and speed. It has almost linear time complexity~\cite{fowler2012towards, higgott2023sparse, wu2023fusion}, is practical for hardware implementation for real-time decoding~\cite{vittal2023astrea} and is more accurate than LUT and Union Find decoders. 

\mwpm constructs a decoding graph from the code as in Fig.~\ref{fig:decoding_graph}. Each node corresponds to a syndrome qubit. Each edge represents an error that can cause the syndrome to flip on its connected nodes. Most edges have two nodes, and some connect one node to the boundary.
In reality, the syndrome extraction circuit and the syndrome measurements also have errors, so the decoder combines multiple rounds of the syndrome (typically $d$). This is referred to as the circuit-level noise model, where the decoding graph is a more complicated 3D graph with edges connecting nodes between adjacent rounds. 

Each edge also has a weight $w=-\log{\frac{p}{1-p}}$. The error probability is obtained from a noise model. In \mwpm assumption, all errors happen independently, and each error will only cause one or two detection events corresponding to one edge in the matching graph. Thus, the highest probability error event can be found as the minimum weight matching of the decoding graph.
However, on real quantum devices, not all errors happen independently, and in many realistic QEC codes, each error may cause more than two detection events. For example, for surface code, a Y error on the data qubit will flip the four syndrome qubits around it, creating four detection events. Besides, since the same syndrome can be caused by different error events, a precise alignment between the decoder's input noise model and real hardware's noise characteristics is crucial for accurate QEC decoding.

\subsection{Decoders Tackling Noise Drift and Correlated Noise}
Several existing works address the issue of noise drift, as detailed in sources like \cite{PhysRevX.13.031007, spitz2018adaptive, kelly2016scalable, wagner2022pauli, wagner2022learning, wagner2021optimal, orsucci2016estimation, fujiwara2014instantaneous, 9838702, 9923816, etxezarreta2021time, PhysRevResearch.5.033055}. Specifically, \cite{fowler2014scalable, fujiwara2014instantaneous, Huo_2017} suggest extracting the noise model directly from decoded surface code errors. \cite{spitz2018adaptive, kelly2016scalable, 9838702} propose improving gate control parameters using the error detection rate of repetition code. \cite{wagner2022pauli, wagner2022learning, wagner2021optimal} demonstrate that the Pauli channel error rate can be accurately estimated from error correction syndromes. However, none of these works address tracking complex correlated errors of multiple qubits using a correlation matrix in DGR. \cite{orsucci2016estimation} investigates recovering the error channel of measurement-based quantum computation from measurement results.

Q3DE~\cite{9923816} proposes enhancing tolerance to multi-bit burst errors (MBBEs) caused by cosmic rays with minimal changes and overhead. DGR focuses on noise drifting, while Q3DE addresses instant transient errors. Q3DE adjusts weights only after detecting an anomalous event, whereas DGR continuously updates edge weights. When Q3DE detects higher error rates due to cosmic rays, it changes the code distance through temporal code expansion, unlike DGR which maintains the same code distance. Additionally, Q3DE does not address correlated errors.

\cite{etxezarreta2021time, PhysRevResearch.5.033055} recommend using time-varying quantum channels (TVQCs) that model fluctuations in relaxation (T1) and dephasing times (T2), rather than static error models, to enhance QEC performance.

Other studies, such as \cite{nickerson2019analysing, fowler2013optimal, criger2018multi, wootton2012high, duclos2010fast, delfosse2014decoding}, address noise correlation issues by altering decoding graph weights. However, these only consider correlations between X/Z errors. Our method handles more complex correlations, like simultaneous X errors on two qubits. Another study~\cite{nickerson2019analysing} modifies graph weights based on syndromes \textit{without} decoding, effectively addressing only local correlations due to the local error diffusive process.

\section{Motivation}

\begin{figure}[t]
    \centering
    \includegraphics[width=\columnwidth]{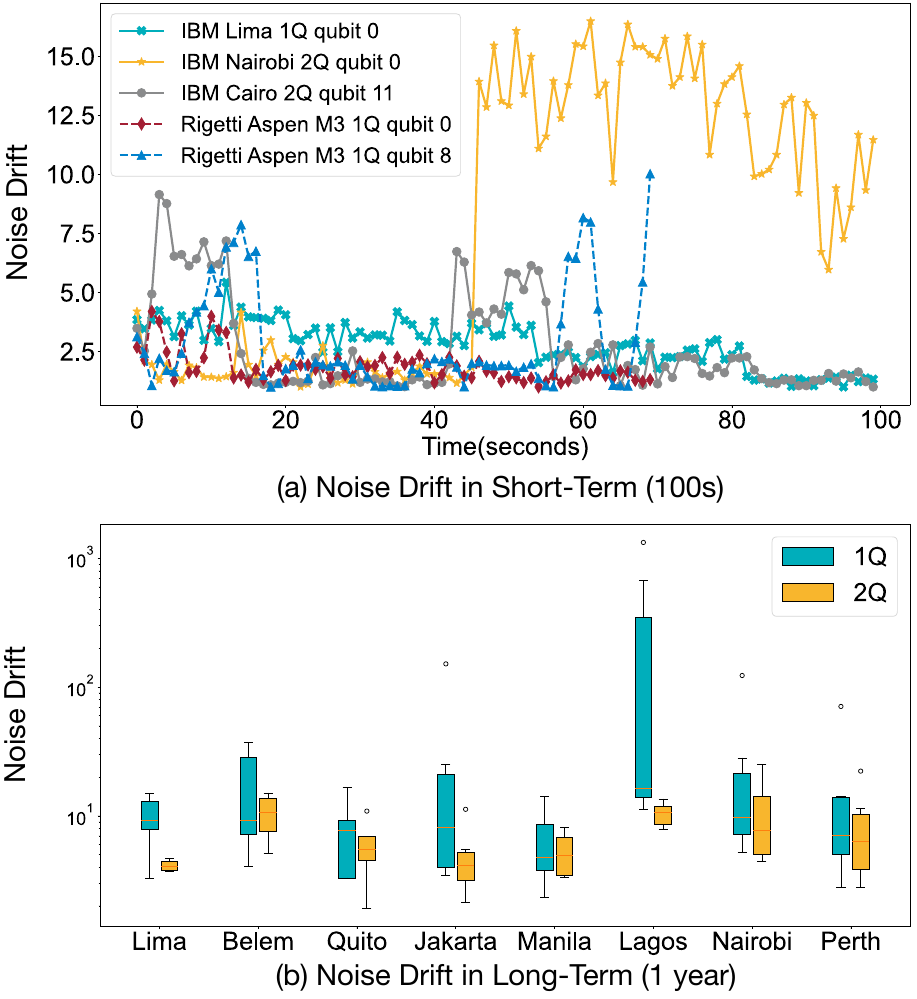}
    \caption{Noise drift with in short-term and long-term.}
    \label{fig:drift}
\end{figure}

\subsection{Noise Drift}

The effectiveness of the decoder is inherently linked to the underlying noise model of the quantum device. Noise models characterize the physical processes that can cause errors, such as bit flips and phase flips, and they provide essential information to compute the probability of particular error patterns. 

Unfortunately, noise in quantum systems is known to undergo significant drift over time~\cite{ravi2023navigating}, which can arise from various sources. 
For superconducting quantum devices, thermal fluctuations can be caused by instantaneous heat released by a quantum phase slip~\cite{gumucs2022calorimetry} in Josephson Junction, leading to noise drift. Unwanted coupling between the two-level-systems (TLS)~\cite{muller2019towards} defects and Transmon qubits~\cite{burnett2019decoherence, schlor2019correlating} can cause the fluctuation of T1 and T2 coherence times. The population of unpaired electrons can also lead to decoherence time changes~\cite{gustavsson2016suppressing}. In neutral atom and trapped-ion systems, photons can be excited from electric-field noise, varying the gate fidelity~\cite{bermudez2017assessing}. Laser instabilities such as detuning~\cite{day2022limits} can also drift the fidelity of gates. Besides, the magnetic field of the Earth is subject to fluctuations influenced by solar phenomena, including the incremental changes that follow the Sun's daily patterns and the more abrupt alterations stemming from events like coronal mass ejections. The variations in the magnetic field can cause fluctuations in qubit frequencies~\cite{mills2022sun}. Cross-talk between qubits, quasi-particles~\cite{klimov2018fluctuations}, and seismic noise from human activities are also responsible for noise drift.

Fig.~\ref{fig:drift} shows our observations on noise fluctuation in IBM and Regetti superconducting devices. Subfig (a) shows the drift of 1Q and 2Q gate error rates within a short period -- 100 seconds. The y-axis is the error rate ratio normalized by the smallest error rate. The drift can exceed 15\x within only 100 seconds. Subfig (b) shows the historical error rate data extracted from IBM devices. We show the distribution of the max drift of the error rate of a gate in the past year. The 1Q and 2Q gate drifts can easily exceed 10\x. 1Q gate drift is larger than 2Q gate, and can be over 1000\x.

To accurately measure and characterize the noise profile, techniques such as quantum process tomography~\cite{chuang1997prescription, PhysRevLett.78.390}, gate set tomography~\cite{nielsen2021gate, greenbaum2015introduction}, randomized benchmarking~\cite{emerson2005scalable, knill2008randomized, magesan2011scalable}, and cross-entropy benchmarking~\cite{boixo2018characterizing} can be employed. The accuracy comes at the cost of inefficiency. These techniques are simply not scalable as the quantum systems grow in complexity. The scalability issue prevents them from being used for noise models that update in real-time. Besides, the techniques above are also impractical to be performed together with the QEC. 
Although some prior work proposed to capture the Pauli error~\cite{wagner2022pauli} from syndromes, we pursue to cover \textit{arbitrary} type of errors (Pauli, measurement, etc.) and capture the \textit{correlations} between different errors.

\begin{figure}[t]
    \centering
    \includegraphics[width=\columnwidth]{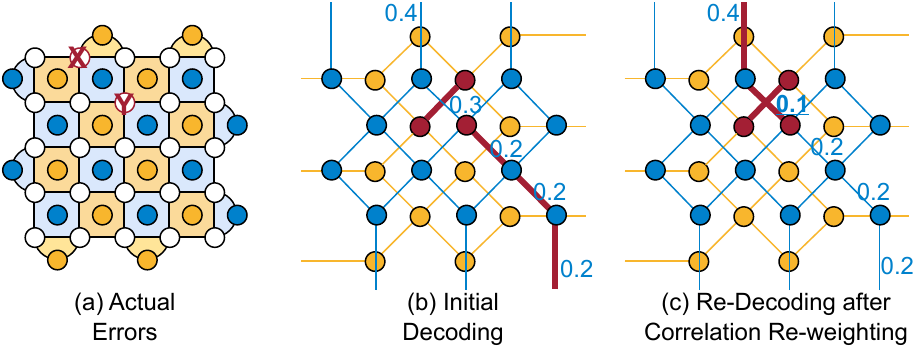}
    \caption{Re-decoding with correlation re-weighting to identify the correct errors.}
    \label{fig:corr_reweighter_example}
\end{figure}

\begin{figure*}[t]
    \centering
    \includegraphics[width=\textwidth]{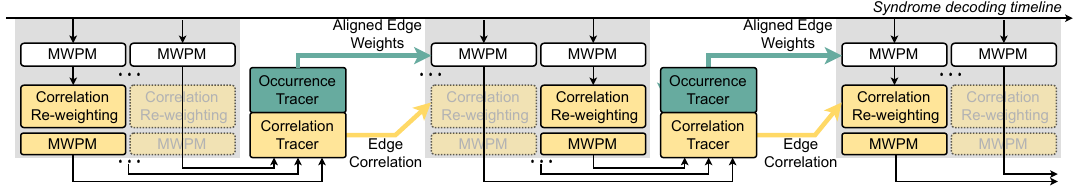}
    \caption{The overview of the proposed decoding graph re-weighting flow, including conditional correlation re-weighting to capture edge correlation and periodic alignment re-weighting to align the probability.}
    \label{fig:flow}
\end{figure*}

\begin{figure*}[t]
    \centering
    \includegraphics[width=\textwidth]{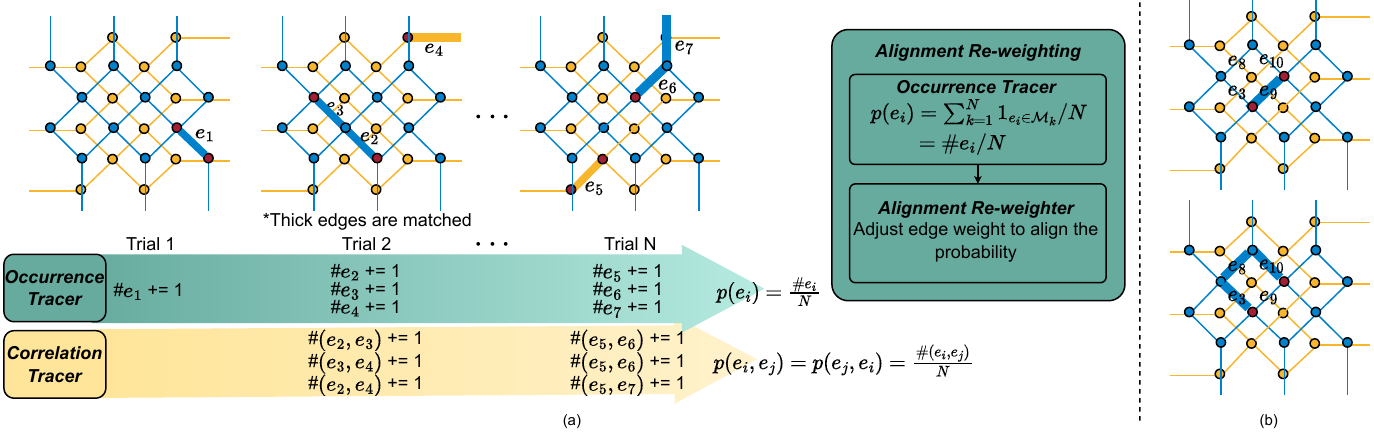}
    \caption{Proposed alignment re-weighting method.
    Based on multiple matching trials, the occurrence tracer and correlation tracer will count the single-edge and edge-pair probability, respectively.
    Based on the occurrence tracing result, the alignment re-weighter will assign the negative log probability as the edge weight.
    The correlation tracer is used in the correlation re-weighting module.}
    \label{fig:align_reweight}
\end{figure*}

\subsection{Noise Correlation}
The \mwpm decoder has the assumption of independence between all the edges in the decoding graph. However, in reality, the edges are correlated. The edge correlations can come from two sources. In the one-to-many type of correlation, one single error can impact multiple edges due to the code construction. For instance, the Y error causes errors on one X and Z edge in surface code; an error on the X-basis measurement of a syndrome qubit causes errors on one Y edge and one Z edge in honeycomb. The second is the many-to-many type of correlation, in which multiple errors happen together and cause correlation in their edges. For example, the 2Q depolarizing channel can cause two Pauli errors on two qubits. Besides, for neutral atom QEC~\cite{bluvstein2022quantum}, one laser controls a whole row or column of qubits, whose error can also be correlated.

The correlation information can be leveraged to augment the \mwpm decoder. For example, in Fig.~\ref{fig:corr_reweighter_example}, the actual errors in (a) contain one X and one Y, which cause three syndromes in (b). With the initial decoding round, the X error matching (blue) will match the syndrome with the bottom boundary because of the lower weight. There are two syndromes for Z error matching (yellow), and the results indicate a Z error. If we know a Y error is causing that Z, we can lower the weight of the corresponding X error edge. The re-decoding process runs \mwpm again and can match the syndrome with the top boundary, which is the correct decoding. In reality, the edge correlations can be obtained from decoding history. 
Some rule-based or data-driven strategies can be developed to perform the re-weighting based on initial matching and correlations.

\section{Methodology}

An overview of \name workflow is shown in Fig.~\ref{fig:flow}. It augments the conventional decoding pipeline by performing alignment re-weighting and correlation re-weighting. The process of alignment re-weighting within the quantum error correction system may be executed either periodically or as necessitated by the drift of noise. It will look back to extract a specific number of immediate preceding decoding trials and leverage the decoded matching from those trials to compute updated weights. These recalculated weights, being in alignment with the existing real noise on the quantum hardware, are then implemented in subsequent \mwpm decoding processes. Moreover, an additional re-weighting step, correlation re-weighting, is strategically employed on what is characterized as ``difficult cases'' to find a more probable error path when taking into consideration the correlation of edges within the system.

\subsection{Alignment Re-weighting}
\label{sec:align_reweighting}
The alignment re-weighting is shown in Fig.~\ref{fig:align_reweight} (a). It contains an occurrence tracer and a re-weighter. The occurrence tracer keeps track of the previous decoding results. When alignment re-weighting is needed, the occurrence tracer estimates the occurrence frequency of each edge in previous rounds and uses this frequency as the probability to update the weight in the MWPM decoding graph. In the figure, we show specific examples. The red circles mark the syndrome locations. Thick edges are the matched edges generated by the \mwpm decoder. The tracer looks back at a set number of decoding trials and keeps a counter for each edge. The counter goes up by one every time that edge appears in the matching.

It is clear that the accuracy of the estimated weight is bounded by the decoder's performance. A poor decoder cannot provide reliable decoder matchings. However, our sensitivity analysis in Section~\ref{sec:evaluation} proves that the re-weighting still works even if the actual error rate drifts by 100\x. Our key insight here is that most syndromes are isolated when the physical error rate is below a certain threshold. For those simple isolated syndromes, the decoding accuracy is robust to the changes in edge weights. Let's consider two syndromes caused by a single, isolated error, as shown in Fig.~\ref{fig:align_reweight} (b). The most straightforward matching is directly matching those two syndromes with one single edge $e_3$. The second shortest path between the syndromes has three edges (another three edges of a square $e_8, e_9, e_{10}$). Since the weight $w$ is the negative logarithm of the probability $p$ (expressed as $w\approx-log(p)$), the error rate $p$ would have to drift roughly three orders of magnitude to cause a mistake in matching. Therefore, our method can handle a wide range of noise drift. If the error is not isolated, the chance of one error in a fixed area is $O(p_{phys})$, and more than two errors is $O(p_{phys}^2)$. As long as $p_{phys}^2\ll p_{phys}$, we can estimate the error with high accuracy and robustness.

\begin{figure}[t]
    \centering
    \includegraphics[width=\columnwidth]{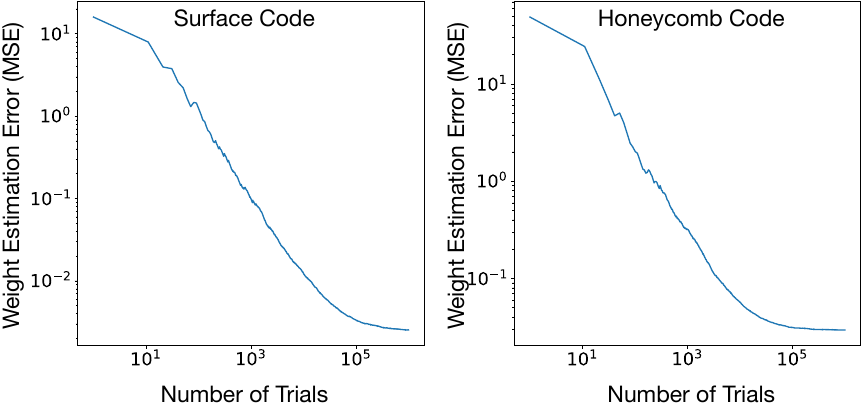}
    \caption{Weight estimation error for surface code with distance $5$ and honeycomb code with distance $3$ under circuit level noise $p=0.001$.}
    \label{fig:weight_error}
\end{figure}

\begin{figure}[t]
    \centering
    \includegraphics[width=\columnwidth]{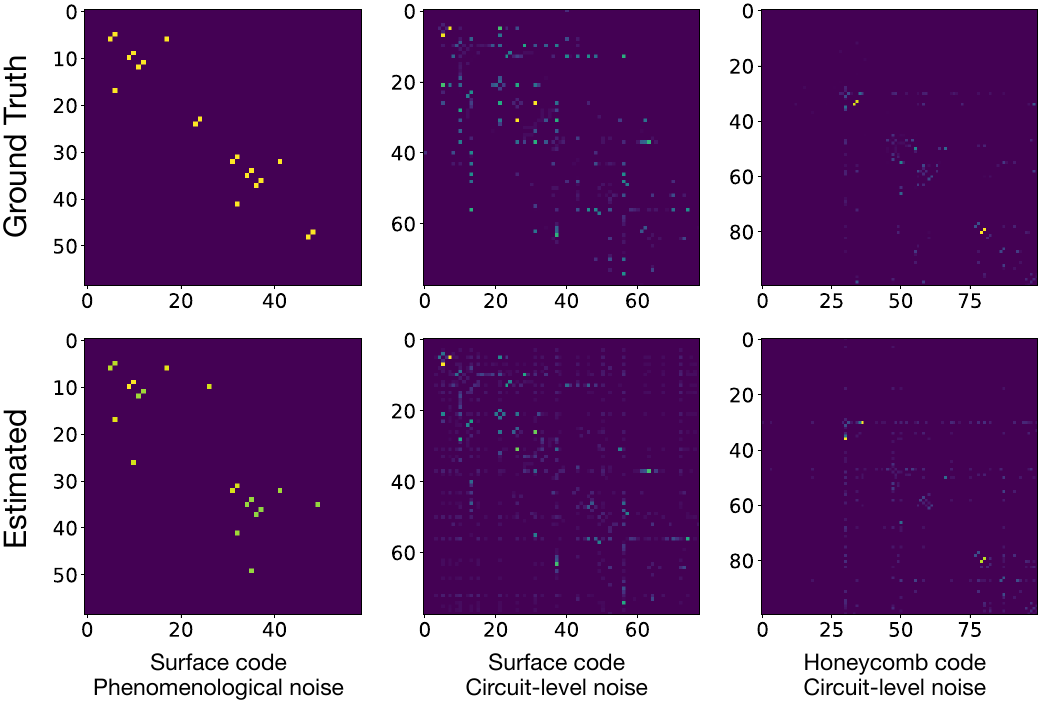}
    \caption{Heat map of the correlation strength for surface code with distance 3 under phenomenlogical and circuit level noise and honeycomb code with distance 3 under circuit level noise. The lighter the point is, the higher the correlation is.}
    \label{fig:heatmap}
\end{figure}
The number of trials required to statistically estimate the weight is influenced by the physical error rate; it increases as the error rate decreases but remains independent of the code distance. 
As illustrated in Fig.~\ref{fig:weight_error}, the mean squared error of estimation tends to converge after $10^6$ trials for both surface code with distance $d=5$ and error rate $p=0.001$, and honeycomb code with distance $d=3$ and error rate $p=0.001$. 

\begin{figure}[t]
    \centering
    \includegraphics[width=\columnwidth]{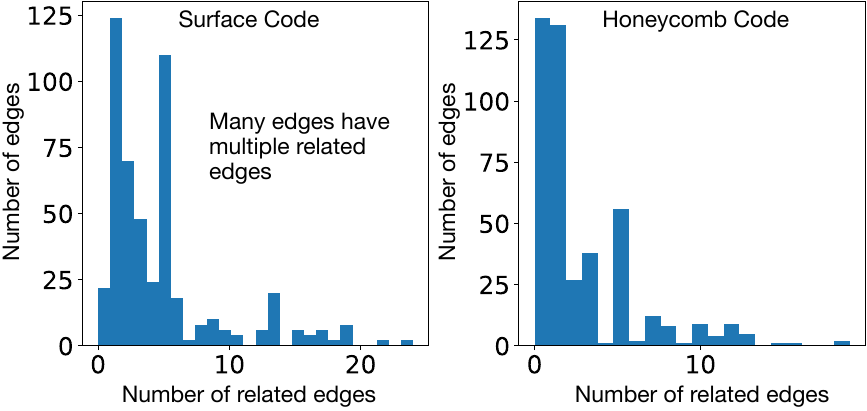}
    \caption{Number of related edges for two codes.}
    \label{fig:n_related_edges}
\end{figure}

\subsection{Correlation Re-weighting}
The second part of \name is called correlation re-weighting. It is comprised of two components: a correlation tracer and a correlation re-weighter. The tracer's function is to construct a correlation matrix by counting the occurrence of edge pairs in previous trials. Though actual correlations may involve multiple edges, here, we restrict our count to correlations solely between two edges for two principal reasons. First, the memory consumption of tracking correlations among multiple edges escalates exponentially. Second, estimating multi-edge correlation is challenging and sometimes impossible. Take the surface code as an example. Suppose the X-check $X_1 X_2 X_3 X_4$ forms a stabilizer. Then, the error $X_1$ and error $X_2X_3X_4$ will generate exactly the same syndromes and can never be distinguished through the measurement data. The ground truth correlation matrices and the estimated ones for surface code and honeycomb code are shown in Fig.~\ref{fig:heatmap}. Most entries in the matrices are accurately estimated, with some discrepancies stemming from the complexities of multi-edge correlations. 

\begin{figure*}[t]
    \centering
    \includegraphics[width=\textwidth]{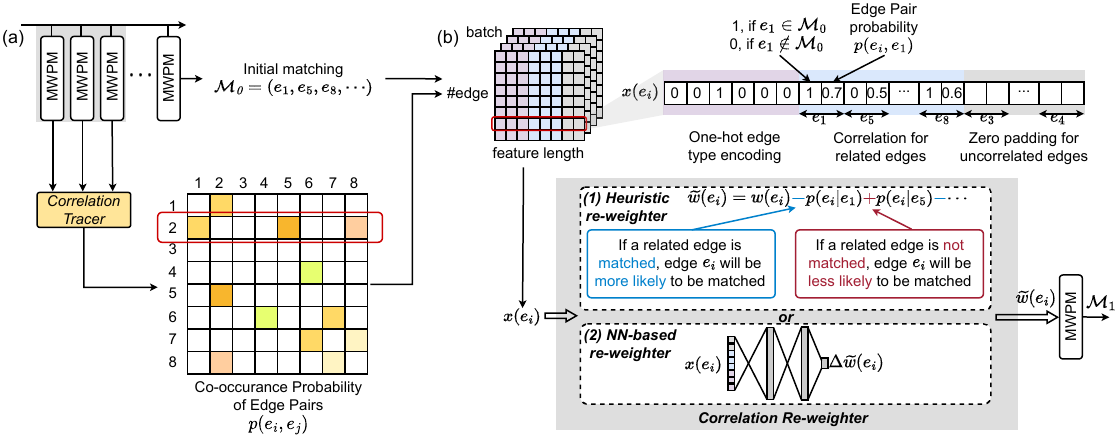}
    \caption{Proposed correlation re-weighting method.
    Once the correlation re-weighting condition is triggered, the edge co-occurrence probability will be obtained from the correlation tracer and fed to the correlation re-weighter, either a heuristic or a NN-based re-weighter, for edge weight adjustment. And then an additional \mwpm is performed with adjusted weights.
    }
    \label{fig:corr_reweight}
\end{figure*}

The MWPM decoder operates by performing certain approximations and views all the edges independently, an approach that can impair its performance. To enhance the decoding process by leveraging correlation information, we introduce a two-iteration decoding methodology. In the first iteration, standard \mwpm decoding is executed. In the second, we adjust the weights based on both the correlation matrix and the initial matching and re-decode the syndromes. The key insight here is that the known occurrence of one edge increases the probability of occurrence for its correlated edges. To limit the overhead for two-iteration decoding, this technique is selectively applied to challenging syndrome patterns (characterized by the number of syndromes), which is inspired by the two-level decoding approach in Clique~\cite{ravi2022have}. The comprehensive illustration of the entire pipeline is shown in Fig.~\ref{fig:corr_reweight}.

\begin{figure}
    \centering
    \includegraphics[width=\columnwidth]{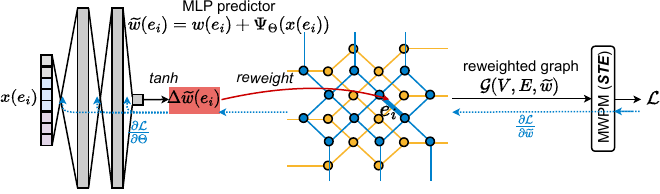}
    \caption{The proposed neural network-based correlation re-weighter trained with SPSA algorithm.
    }
    \label{fig:nn_predictor}
\end{figure}

\subsubsection{Heuristic-Based Correlation Re-weighter}
The core intuition behind correlation re-weighting lies in the relationship between physical errors and their corresponding edges within the matching graph. Specifically, if a single physical error leads to two or more edges in this graph, a positive correlation emerges among these edges. Consequently, if one edge is chosen in the predicted matching, the probability of the other edges occurring should be \textit{increased}, resulting in a decrease in their weights. Conversely, if one edge is not selected, the probability of the occurrence of the other correlated edges should be \textit{diminished}, leading to an increase in weights.

With regard to the correlation re-weighter, one straightforward heuristic is to re-weight according to the conditional probability of edge matching in the initial matching~\cite{demarti2023performance}. Unfortunately, this approach is viable only when an edge has one single correlated edge. However, as in Fig.~\ref{fig:n_related_edges}, after evaluating the number of related edges on surface code with distance 5 and honeycomb code with distance 3, it's apparent that the majority of edges have more than one correlated edge, so the application of a simple conditional probability becomes impossible. This complexity necessitates a more nuanced approach to accurately capture the correlations and their effects on the decoding process.
As a result, we propose a heuristic re-weighting algorithm that is able to process multiple correlated edges, increasing or decreasing the edges' weight according to the strength of correlation.

After the first decoding round, we get the initially selected edge set $M$; then, we update all the edges' weights according to the following rule.
\begin{equation}
\begin{split}
    \widetilde{w_j}&=w_j-\sum_{e_i\in M}\frac{p(e_i,e_j)}{p(e_i)}+\sum_{e_i\not \in M}\frac{p(e_i,e_j)}{p(e_i)} \\
        \label{eq:he_update_rule}
\end{split}
\end{equation}
where $w_j$ and $p(e_i)$ represent the weight and occurrence probability of edge $e_j$, $e_i$, respectively, as estimated by the occurrence tracer, and $p(e_i,e_j) $ denotes the probability of edge $e_i$ and $e_j$ co-occuring. Subsequently, we proceed with a second decoding round, utilizing these newly calculated weights to refine the results.

\begin{algorithm}[t]
\caption{Hybrid training for NN reweighing predictor}
\label{algo:mlptrain}
\renewcommand{\algorithmicensure}{\textbf{Initialize:}}
\begin{algorithmic}[1]
\ENSURE{Decoding graph G, NN predictor $\Psi_{\Theta}(\cdot)$, Dataset $\mathcal{D}$, Steps $T$, Learning rate $\eta$, initial weight $w$}
\FOR{each training iteration $t\gets 1\cdots T$}
\STATE Sample a mini-batch $(x,m')$ from $\mathcal{D}$
\STATE \textcolor{gray}{Forward the MLP predictor}
\STATE $\widetilde{w}= w + \Psi_{\Theta^t}(x)$ 
\STATE \textcolor{gray}{Forward the MWPM decoder and calculate the loss}
\STATE $\mathcal{L}(\widetilde{w})=\mathcal{L}(\texttt{MWPM}(G(\widetilde{w})), m')$ 
\STATE \textcolor{gray}{Estimate the gradient through the combinatorial MWPM decoder}
\STATE $\nabla_{\widetilde{w}}\mathcal{L}=\frac{1}{2Q\sigma^2}\sum_{i=1}^Q\big(\mathcal{L}(\widetilde{w}+\Delta w_i)-\mathcal{L}(\widetilde{w}-\Delta w_i)\big)\Delta w_i$
\STATE \textcolor{gray}{Use the estimated gradient for backpropagation}
\STATE $\nabla_{\Theta^t}\mathcal{L}=\nabla_{\widetilde{w}}\mathcal{L}\cdot\nabla_{\Theta^t}\widetilde{w}$
\STATE \textcolor{gray}{Update the NN predictor weights}
\STATE $\Theta^{t+1}\gets\Theta^t-\eta \nabla_{\Theta}\mathcal{L}$

\ENDFOR
\RETURN{Trained NN predictor $\Psi_{\Theta^*}(\cdot)$} \\
\end{algorithmic}
\end{algorithm}

\subsubsection{NN-based Correlation Re-weighter}

\begin{figure*}[h]
    \centering
    \includegraphics[width=\textwidth]{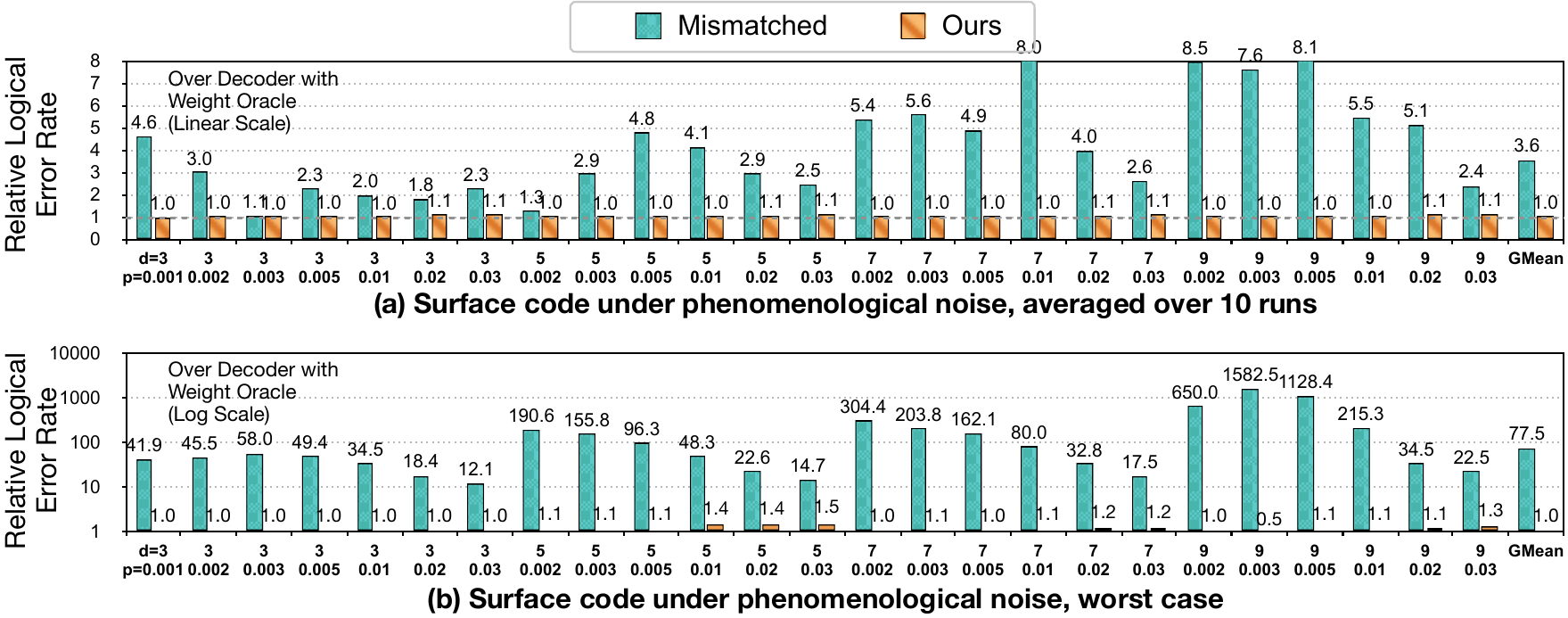}
    \caption{Logical error rate with and without alignment re-weighting on 10\x mismatch for surface code under phenomenological noise.}
    \label{fig:ler_w_wo_prediction_surface_pheno}
\end{figure*}

If the edge is associated with a multitude of related edges with complex relationships, the heuristic-based re-weighting may not provide effective results.
To fully leverage the correlation matrix derived from the matching history, we introduce a neural network-based re-weighter. This re-weighter is designed to predict weights changes, denoted as $\Delta\widetilde{w}(e_i)=\Psi_{\Theta}(x(e_i))$, where $x(e_i)$ symbolizes the extracted feature for edge $e_i$, as shown in Fig.~\ref{fig:corr_reweight} and Fig.~\ref{fig:nn_predictor}, and $\Psi_{\Theta}(\cdot)$ is the NN-based predictor. The weight for the second iteration of decoding is calculated as $\widetilde{w}(e_i) = w(e_i) + \Delta\widetilde{w}(e_i)$. 
The extracted feature contains information on the edge type, which is encoded as a one-hot vector. The edge type is determined by the spatial location of an edge in the decoder graph. The edges of the same type will have the same number of correlated edges. Furthermore, the feature vector incorporates information on correlated edges, including whether the related edge has been matched in the initial iteration matching $M_0$ and the co-occurrence probability $p(e_i,e_j)$. We make sure that the order of the correlated edges in the vector consistently follows the same related location order for one edge type, such as from front to back, from top to bottom, etc. This uniform ordering is necessary to maintain the spatial invariance of the code. To accommodate varying numbers of related edges, the feature vector is zero-padded to the maximum number of related edges, ensuring a consistent input for training and inference.

The complexity of model training represents the most demanding part of this process, and it can be mathematically formulated as
\begin{equation}
    \min_{\Theta}~~\sum_{e_i\sim E}\mathcal{L}\Big(\texttt{MWPM}\big(G(w+\Psi_{\Theta}(x(e_i)))\big), m'\Big),
\end{equation}
where the objective is to minimize the error between the ground-truth matching solution $m'$ and the solution on the re-weighted decoding graph with the predicted edge changes of the weights $\Delta w$.

The gradients w.r.t. the NN weights are expressed as$\nabla_{\Theta}\mathcal{L}=\nabla_{\widetilde{w}}\mathcal{L}\cdot\nabla_{\Theta}\widetilde{w}$.
A significant challenge arises from the combinatorial nature of the solver, \texttt{MWPM}($\cdot$), which is not differentiable, i.e., the first term $\nabla_{\widetilde{w}}\mathcal{L}$ cannot be obtained directly.
To navigate around this limitation and incorporate the black-box combinatorial solver into the backpropagation procedure, we adopt a symmetric zeroth-order gradient estimator to obtain the directional derivative, as described in Alg.~\ref{algo:mlptrain} Lines 7-10.
To achieve this, we randomly generate a perturbation of the weights from a multi-variate Gaussian distribution $\Delta w\sim \mathcal{N}(0,\sigma^2)$, and use the directional derivative as an estimate of the first-order gradient of the smoothed \mwpm oracle.
To reduce the sampling variance, we average the gradients over $Q$ samples.
Subsequently, the estimated gradients of the solver are used to calculate the gradients of the weights of the NN model.

During the inference phase, the model can process all edges in a batch, thereby predicting the $\Delta\widetilde{w}(e_i)$ to minimize latency.

\section{Evaluation}

\label{sec:evaluation}
\subsection{Evaluation Methodology}
\noindent\textbf{Benchmarks.} We evaluate two quantum error correction codes, the surface and honeycomb codes. For surface code with code distance $d$, we use \texttt{stim}'s~\cite{gidney2021stim} built-in functions \texttt{surface\_code:rotated\_memory\_z} to generate a surface code with $d$ rounds and test with logical Z observable. For honeycomb code, we use the package provided by Ref.~\cite{gidney2022benchmarking} to generate a planar honeycomb code. For a code distance with $d$, we use $d+1$ width, $3 \times [(2d + 2) / 3] $ height and $3d$ rounds, and test with a V-type observable. We use PyMatching~\cite{higgott2023sparse} to sample the errors and decode them. We select distances 3, 5, 7, and 9 for the surface code with physical error rates from 2E-4 to 3E-2. We select distances 3 and 5 for the honeycomb code with a physical error rate from 3E-5 to 1E-3.

\begin{figure*}[t]
    \centering
    \includegraphics[width=\textwidth]{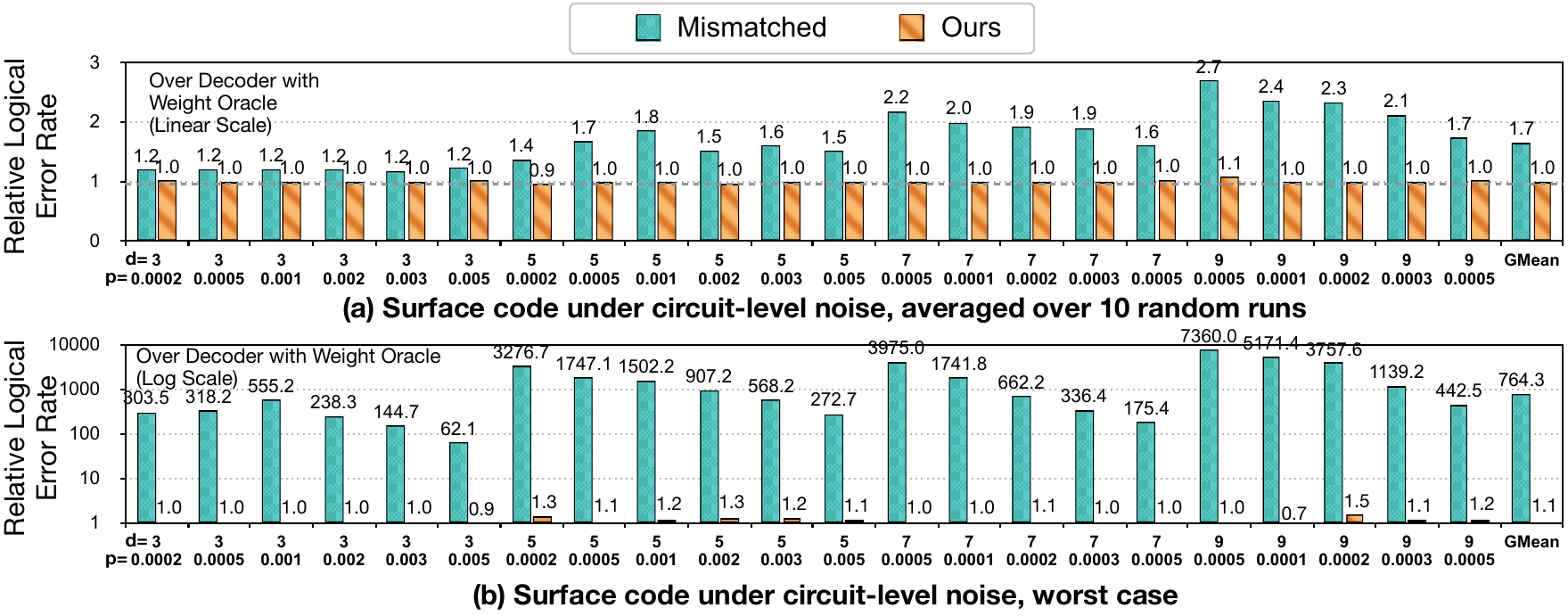}
    \caption{Logical error rate with and without alignment re-weighting on 10\x mismatch for surface code under circuit-level noise.}
    \label{fig:ler_w_wo_prediction_surface_circuit}
\end{figure*}

\noindent\textbf{Noise Model.} We implement two types of noise model, the phenomenological noise model~\cite{dennis2002topological}, which contains data qubit depolarization and measurement error, and the circuit-level noise model, which includes data qubit depolarization, depolarization after Clifford gates, and measurement error. For surface code, the errors are generated using \texttt{stim}'s built-in functions, and for honeycomb code, we used the SD6 circuit level noise model introduced in~\cite{gidney2022benchmarking}.

\noindent\textbf{Mismatch.} We evaluate two mismatch models. The first one is random (average-case mismatch). We first generate an error correction circuit with a given physical error rate in the random mismatch model. Then, we mutate each operation's error probability with a factor, sampled log-uniformly from $\frac{1}{N}$ to $N$. This is called $N$\x mismatch. The circuit and the matching are sampled independently. The second one is the \textit{worst-case mismatch}, in which the difference between the logical error rate using the matched decoder and the mismatched one is maximized. We found that for surface code with logical Z observable, the logical error rate will be maximized, if the first $\frac{d-1}{2}$ rows of the data qubits have a larger physical error rate and the last $\frac{d+1}{2}$ rows have a lower physical error rate.

\noindent\textbf{NN Decoder Settings}
The NN we adopt is an MLP with two hidden layers with dimension 64. The last layer is a \texttt{tanh} output layer scaled by 3\x, outputting $[-3, 3]$. We leverage the Adam optimizer with a learning rate of 1E-3 and weight decay of 1E-4. The models are trained with batch size 128 for 100 epochs on datasets with 100000 generated samples.

\begin{figure}[t]
    \centering
    \includegraphics[width=\columnwidth]{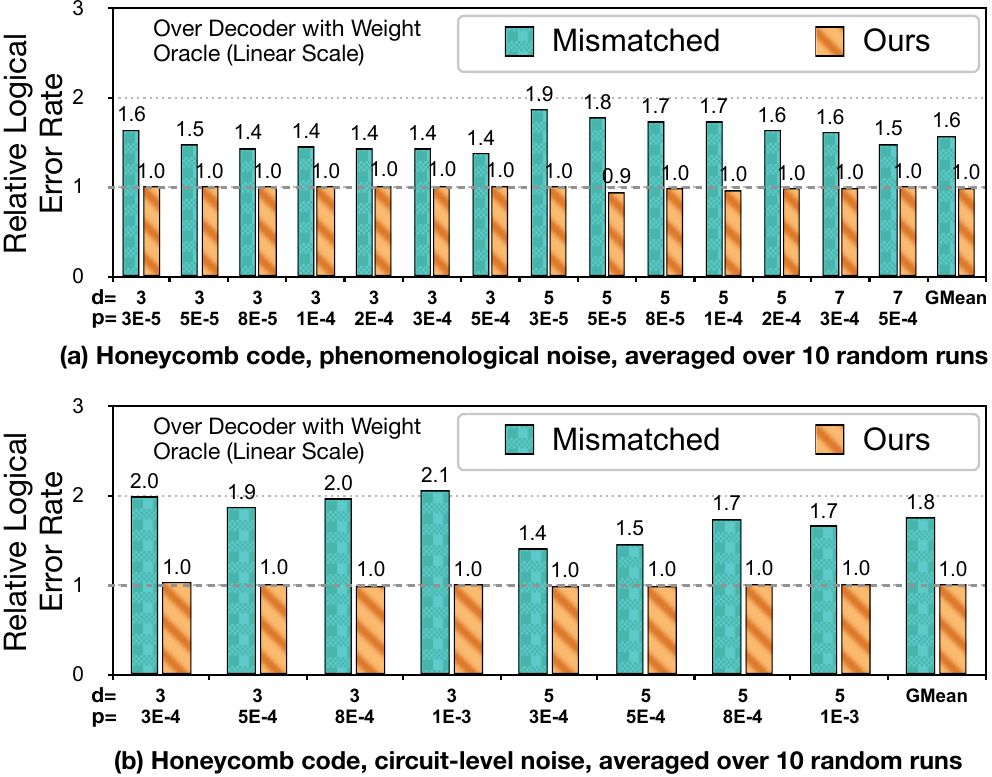}
    \caption{Logical error rate with and without alignment re-weighting on 10\x mismatch for honeycomb code under phenomenological and circuit-level noise.}
    \label{fig:honey_align_bar}
\end{figure}

\begin{figure}[t]
    \centering
    \includegraphics[width=\columnwidth]{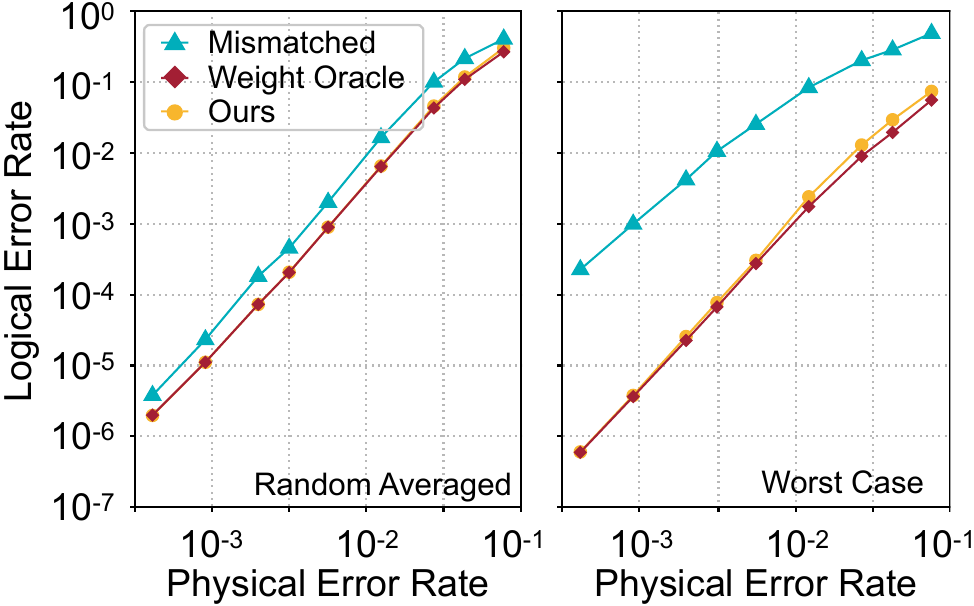}
    \caption{Logical error rate with and without matching prediction on 10\x mismatch.}
    \label{fig:surf_align_curve}
\end{figure}

\begin{table}[t]
\centering
\caption{Physical error rate threshold for surface code.}

\renewcommand*{\arraystretch}{1}
\setlength{\tabcolsep}{10pt}

\resizebox{\columnwidth}{!}{
\begin{tabular}{lccc}
\toprule
Mismatch Strength&Oracle&Mismatched&Ours\\\midrule
10$\times$&0.0301&0.0194&0.0300\\
5$\times$&0.0272&0.0234&0.0271\\
\bottomrule
\end{tabular}
}
\label{tab:threshold}

\end{table}

\begin{figure}[t]
    \centering
    \includegraphics[width=\columnwidth]{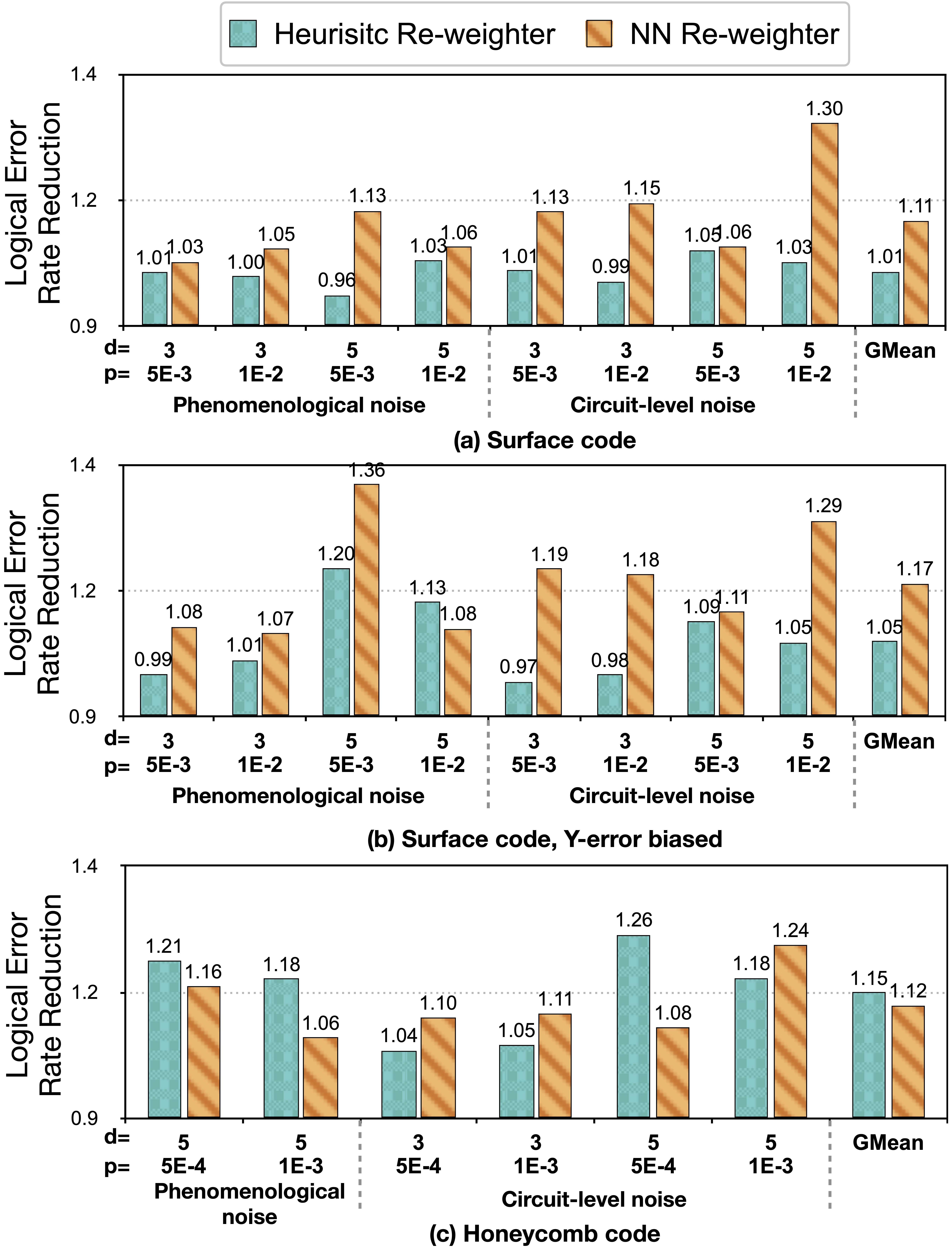}
    \caption{Logical error rate reduction with the proposed correlation re-weighting scheme.}
    \label{fig:corr_bar}
\end{figure}

\subsection{Main Results}

\noindent\textbf{Logical error rates reduction with alignment re-weighting.} Fig.~\ref{fig:ler_w_wo_prediction_surface_pheno} shows the performance of surface code under phenomenological noise. The y-axis is the relative logical error rate of decoder with mismatch weights and our aligned weights, over the decoder with matched weight provided by an oracle. Our aligned weights is obtained from matching of 1 million trials. The mismatch strength is 10\x. Subfig. (a) shows the average performance over 10 random sampled noise model. The mismatched weights degrade the decoder by 3.6\x on average, while our alignment re-weighter can fully recover the logical error rate. Generally, we observe the improvements of our re-weighter is more significant for large distances and low physical error rates. Subfig. (b) shows the performance in the worst case mismatch where the mismatched weights can cause logical error rate to degrade by up to 1582\x, with 77.5\x on average. Fig.~\ref{fig:ler_w_wo_prediction_surface_circuit} further compares the mismatched weight with our aligned weight under the circuit-level noise model. For random benchmarks, \name provides 1.7\x improvements and under worst case mismatch, \name provides 695\x reduction on average (764.3/1.1), up to 7360\x. Our aligned weights cannot fully recover performance, leading to a 1.1\x higher logical error rate. Fig.~\ref{fig:honey_align_bar} provides more results on the honeycomb code. Again, \name reduces logical error rate by 1.6\x and 1.8\x for two noise models and fully recovers them to the level of using weight oracles. 
From these results, we demonstrate the detrimental effect of mismatched weights, and prove the effectiveness of our \name on various code types, distances and error rates. 

 \begin{figure}[t]
    \centering
    \includegraphics[width=\columnwidth]{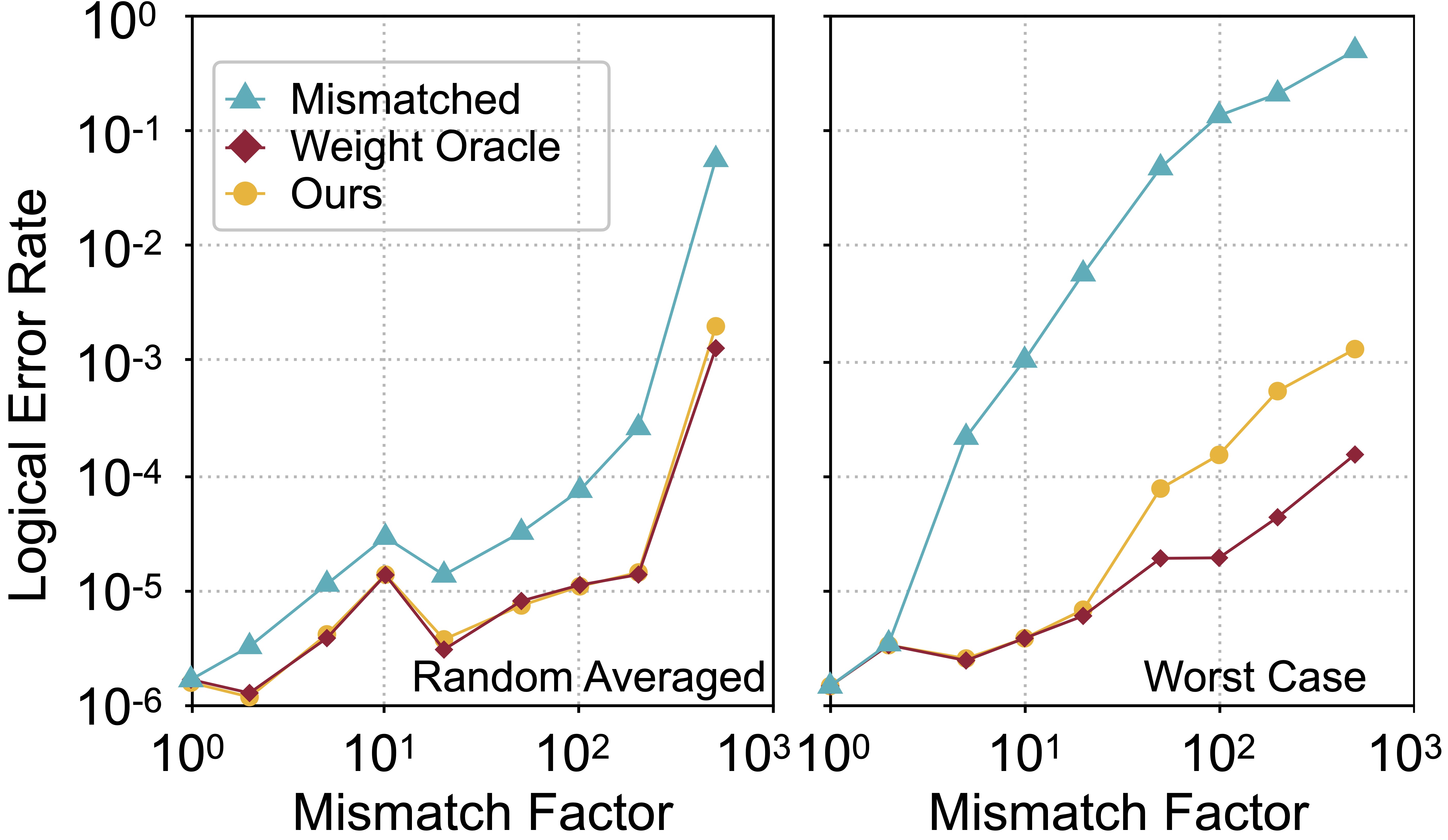}
    \caption{Performance of alignment re-weighter v.s. different mismatch level. Evaluated on surface code with distance 5 under phenomenological noise with $p=0.001$.}
    \label{fig:mismatch_vs_l}
\end{figure}
\begin{figure}[t]
    \centering
    \includegraphics[width=\columnwidth]{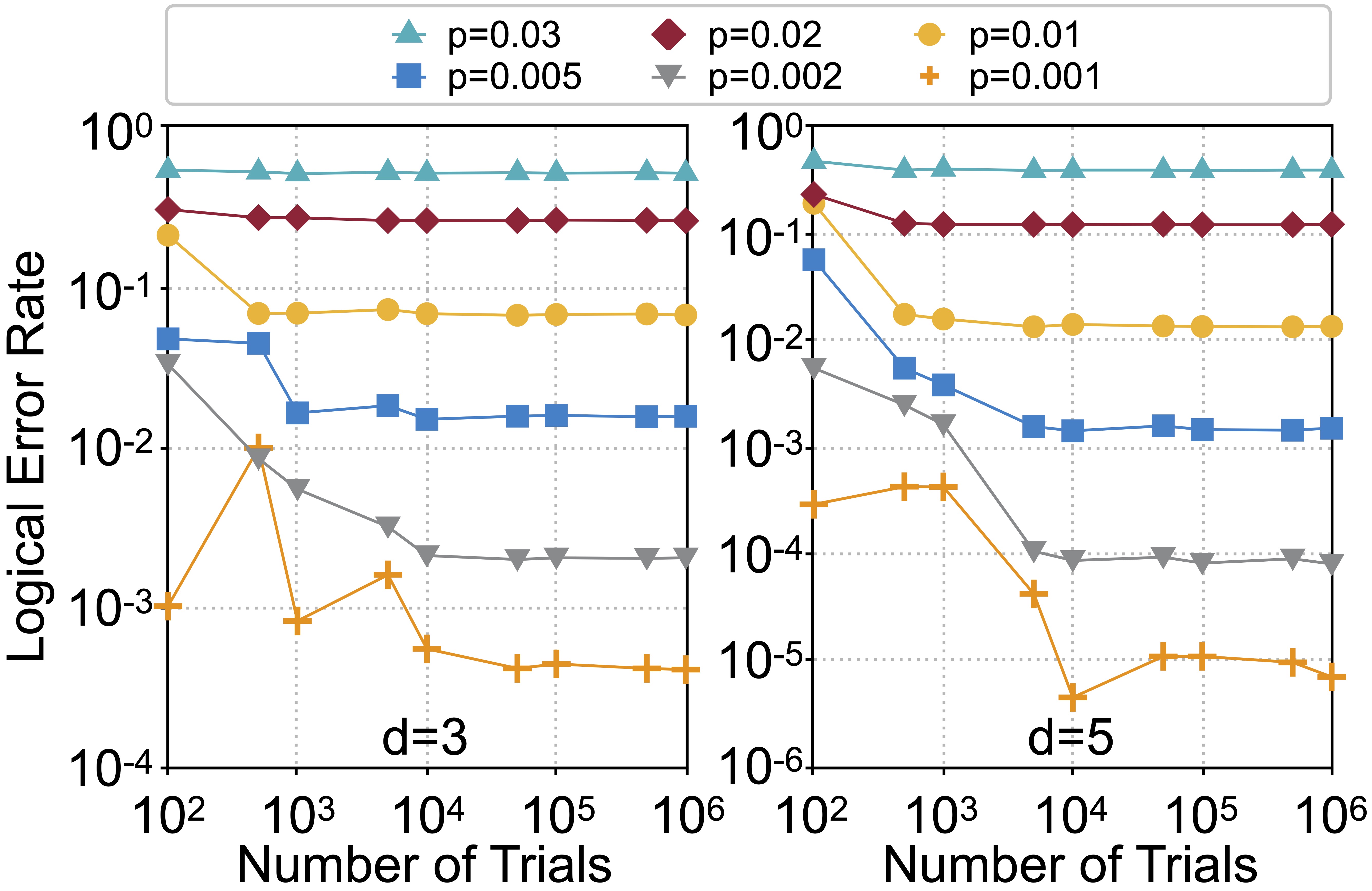}
    \caption{Logical error rate vs. the number of trials under different noise level, evaluated on surface code under phenomenological noise.}
    \label{fig:trail_vs_l_converge}
\end{figure}

\begin{figure*}[t]
    \centering
    \includegraphics[width=\textwidth]{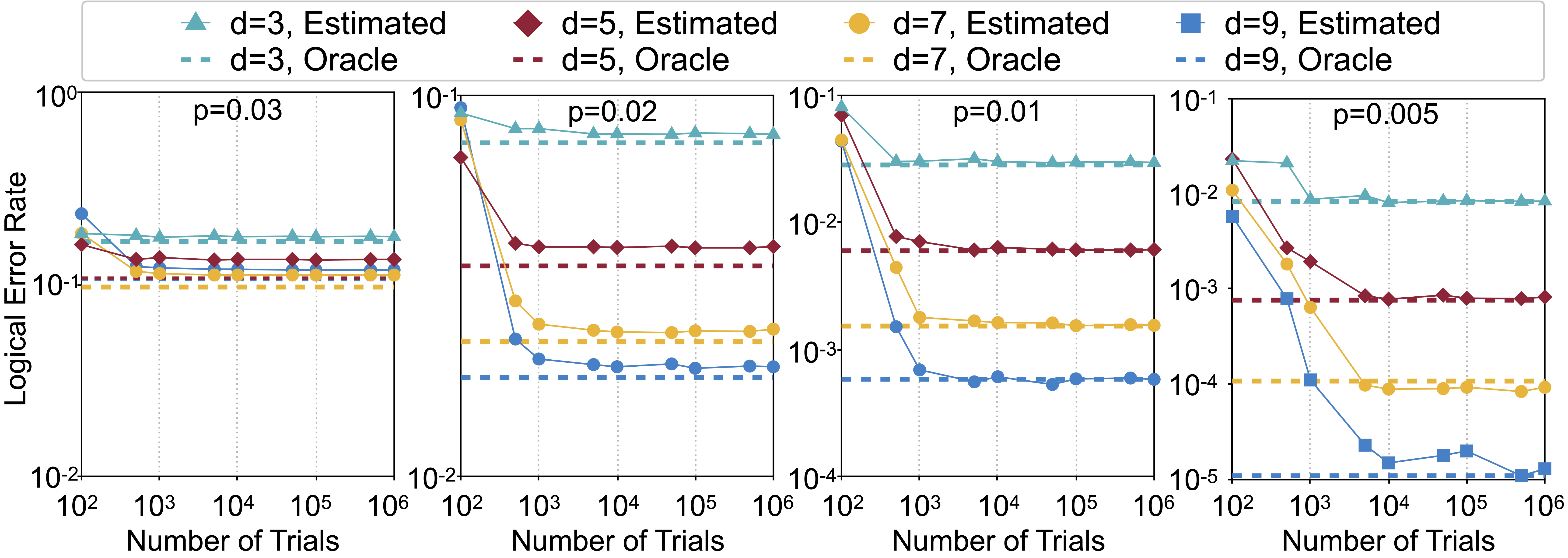}
    \caption{Logical error rate vs. the number of trials with different code distances, evaluated on surface code under phenomenological noise.}
    \label{fig:trail_vs_l}
\end{figure*}

\begin{table}[t]
\centering

\renewcommand*{\arraystretch}{1}
\setlength{\tabcolsep}{7pt}

\caption{Required memory size for storing 10000 rounds of matching trials on classical machine.}
\label{tab:memory}
\resizebox{0.95\columnwidth}{!}{%
\begin{tabular}{lcccccc}
\toprule
Memory (KB) & p=1E-3 & =p2E-3 & p=5E-3 & p=1E-2 & p=2E-2 & p=3E-2 \\
\midrule
d=3&0.39& 0.78& 1.95& 3.9& 7.8& 11.7 \\
d=5&2.51& 5.02& 12.55& 25.1& 50.2& 75.3 \\
d=7&7.79& 15.58& 38.95& 77.9& 155.8& 233.7 \\
d=9&17.67& 35.34& 88.35& 176.7& 353.4& 530.1 \\

\bottomrule
\end{tabular}%
}
\end{table}

\begin{table}[t]
\centering

\renewcommand*{\arraystretch}{1}
\setlength{\tabcolsep}{2pt}

\caption{Number of edges, feature dimension and latency of NN re-weighter for different QEC.}
\label{tab:latency}
\resizebox{0.95\columnwidth}{!}{%
\begin{tabular}{lccccccc}
\toprule
Code & surface & surface  & surface & surface & honeycomb & honeycomb \\
Distance & 3&5&3&5&3&5 \\
Error Model & pheno & pheno & circ & circ & circ & circ \\
\midrule
N\_edges & 59 & 313 & 86 & 518 & 441 & 2499  \\
N\_features & 15 & 18 & 80 & 139 & 148 & 221 \\
Latency (ms) & 0.14 & 0.23 & 0.15 & 0.31 & 0.28 & 0.77\\
\bottomrule
\end{tabular}%
}
\end{table}

\noindent\textbf{Logical error curve.} Fig.~\ref{fig:surf_align_curve} shows the logical error rate on surface code with distance 5 on the phenomenological noise model. As the physical error rate increases, the ratio between the mismatched logical error rate and the matched one decreases. This is because both of them will converge to 0.5 when the physical error rate is large enough. The trend on the worst case mismatch is more significant. When the physical error rate increases, our method's performance slightly decreases because more errors happen in the decoding procedure, thus making the estimation of weights inaccurate. 

\noindent\textbf{Physical error threshold.} Tab.~\ref{tab:threshold} shows the physical error threshold for surface code. In both 10\x and 5\x mismatch strength, the threshold of the mismatched decoder is significantly lower, and the predicted weights have a performance comparable to the ideal weight oracle. 

\noindent\textbf{Logical error rates reduction with correlation re-weighting.} Fig.~\ref{fig:corr_bar} (a) and (b) show the logical error rate reduction for surface code with heuristic and NN correlation re-weighter on two different noise models. The heuristic re-weighter reduces the logical error by 1.01\x and 1.05\x on average for surface code and 10\x Y-biased surface code, while the NN re-weighter manages to reduce by 1.11\x and 1.17\x.
Fig.~\ref{fig:corr_bar} (c) shows the logical error rate reduction for honeycomb code with heuristic and NN re-weighter on the phenomenological and circuit level noise models. The heuristic re-weighter reduces the logical error by 1.15\x on average, while the NN re-weighter reduces the logical error by 1.12\x. Although the heuristic-based re-weighter sometimes works better than the NN-based re-weighter. The NN-based one has a more stable performance and is capable of handling different types of QEC codes under diverse settings.

\subsection{Analysis}
\noindent\textbf{How severe mismatch can our method recover?} Fig.~\ref{fig:mismatch_vs_l} shows the performance of alignment re-weighter when increasing the mismatch level. For the random case, the alignment re-weighter can always recover the correct weight, up to 500\x mismatch. In the worst case, the alignment re-weighter can recover the weight almost perfectly when the mismatch is smaller than 50\x. After that, the logical error rate degrades when the mismatch is more severe, but it is still better than the mismatched decoder. As analyzed in Section~\ref{sec:align_reweighting}, we expect the weight estimation will be less accurate when a direct edge between two syndrome nodes could be mis-decoded to a three-edge detour. That will happen when the mismatch strength is 1000\x, which is close to our observations here for the average-case mismatch.

\noindent\textbf{Required trials for accurate estimation vs. physical error rate.} Fig.~\ref{fig:trail_vs_l_converge} shows the convergence of logical error rates under different noise levels. The smaller the noise is, the more trials are required. This is because whether an edge is selected forms a Poisson distribution with expectation $p$. Estimating the expectation value up to a constant relative error requires trials inversely proportional to $p$.

\noindent\textbf{Required trials for accurate estimation vs. code distance.} Fig.~\ref{fig:trail_vs_l} shows the convergence of logical error rate with the increasing of trials. For different code distances, the required trial is similar, around $10^3$ to $10^4$. This is because, under the same error rate, the error occurrence of an edge is not influenced by the size of the decoding graph.

\noindent\textbf{Memory Requirements.} Table~\ref{tab:memory} shows the additional memory requirement for the occurrence and correlation tracers to work. We use a sparse matrix with INT16 data type to store previous decoding results up to N trials. Thus, whenever needed, the two tracers can estimate the weights and correlations from the data. For the most resource-consuming case, $d=9$ and $p_{phys}=0.03$, the memory requirement is only 0.5MB. The memory requirement can be further decreased by storing more coarse-grained data.

\noindent\textbf{Overhead of NN-reweighter.} Table~\ref{tab:latency} shows the overhead of the NN re-weighter. The latency ranges from 0.14ms to 0.77ms according to the scale of the problem. We can further reduce the average latency by only choosing the difficult cases with more syndromes to go through the NN re-weighter. In our experiments above, only 15\% of trials need correlation re-weighting. The easy cases with fewer syndromes can be correctly decoded in the first iteration with high probability.

\section{Conclusion and Outlook}

We propose \name, an efficient decoding graph edge re-weighting strategy to solve the noise drift-unawareness and correlation-unawareness of the \mwpm decoder. 
\name encapsulates alignment re-weighting that statistically estimates the probabilities of individual edge according to the history of decoded matchings and update the edge weights. Besides, correlation re-weighting takes the correlation between edges into consideration to adjust the edge weight and perform re-decoding. We propose one heuristic-based re-weighter and one neural network-based re-weighter, and both of them can improve the logical error rate leveraging correlations.
Looking forward, we expect the noise drifts and correlations on large-scale quantum systems will likely be more complicated. We hope \name provides a perspective on dealing with them with small classical overhead and no quantum overhead.

\section*{Acknowledgement}
This work is partially supported by MIT-IBM Watson AI Lab and LPS/ARO QCISS program W911NF-21-1-0005.

\bibliographystyle{IEEEtranS}
\bibliography{main, mypaper}

\end{document}